\documentclass[11pt]{article}
\usepackage{amsfonts}
\usepackage{latexsym,amsmath}
\usepackage{amssymb,array}
\usepackage{calc}
\RequirePackage[dvips]{graphicx}
\usepackage{graphics}
\usepackage[colorlinks=true]{hyperref}
\hypersetup{urlcolor=blue, citecolor=red}
\makeatletter \oddsidemargin 0in \evensidemargin 0in \textwidth
16cm\textheight 20cm 
\begin{document}
\newcommand{\la}{\lambda}
\newcommand{\eq}{\Leftrightarrow}
\newcommand{\mf}{\mathbf}
\newcommand{\ri}{\Rightarrow}
\newtheorem{t1}{Theorem}[section]
\newtheorem{d1}{Definition}[section]
\newtheorem{n1}{Notation}[section]
\newtheorem{c1}{Corollary}[section]
\newtheorem{l1}{Lemma}[section]
\newtheorem{r1}{Remark}[section]
\newtheorem{e1}{Counterexample}[section]
\newtheorem{p1}{Proposition}[section]
\newtheorem{cn1}{Conclusion}[section]
\renewcommand{\theequation}{\thesection.\arabic{equation}}
\pagenumbering{arabic}
\title {Stochastic Comparison of Parallel Systems with Log-Lindley Distributed Components under Random Shocks}
\author{Shovan Chowdhury\\Indian Institute of Management, Kozhikode\\Quantitative Methods and Operations Management Area
\\Kerala, India\\E-mail: shovanc@iimk.ac.in \and Amarjit Kundu\footnote{Corresponding
author e-mail:
bapai\_k@yahoo.com}\\Raiganj University\\Department of
Mathematics\\ West
Bengal, India\\Email: bapai\_k@yahoo.com
}
\maketitle
\begin{abstract}
In this paper we compare two parallel systems of heterogeneous-independent log-Lindley distributed components using the concept of matrix majorization. The comparisons are carried out with respect to the usual stochastic ordering when each component receives a random shock. It is proved that for two parallel systems with a common shape parameter vector, the majorized matrix of the scale and shock parameters leads to better system reliability. It is also shown through counter examples that no such results exist when the matrix of shape and shock parameters of one system majorizes the same of the other.
\end{abstract}
\section{Introduction}
\setcounter{equation}{0}
\hspace*{0.3in} Stochastic comparison of system lifetimes has always been a relevant topic in reliability optimization and life testing experiments. If $X_{1:n}\leq X_{2:n}\leq\ldots\leq X_{n:n}$ denote the ordered lifetimes of the random variables $X_1, X_2,\ldots,X_n$, then the lifetime of series and parallel systems correspond to the smallest ($X_{1:n}$), and the largest ($X_{n:n}$) order statistics respectively. Several papers have dealt with comparisons among systems (largely on the parallel and the series) with heterogeneous independent components following a certain probability distribution with unbounded/bounded support, such as exponential, gamma, Weibull, generalized exponential, generalized Weibull, Fr$\acute{e}$chet, beta, Kumaraswamy, or log-Lindley. One may refer to Dykstra \emph{et al.}~\cite{dkr11}, Zhao and Balakrishnan~\cite{zb11.2}, Balakrishnan \emph{et al.}~\cite{ba1}, Fang and Zhang~\cite{fz}, Torrado and Kochar~\cite{tr11}, Kundu \emph{et al.}~\cite{kun1}, Kundu and Chowdhury~\cite{kun2},\cite{kun3}, Gupta \emph{et al.}~\cite{gu}, Chowdhury and Kundu~\cite{ch} and the references therein for more detail. The assumptions in the papers are that the components of the system fail with certainty and the comparison is carried out on the minimums or the maximums of the failed components. Now, it may so happen that the components experience random shocks which may or may not result in the failure of the components. Consider the following example:\\ 
Let us assume a parallel system having $n$ independent components each of which is in working conditions. Each component of the system receives a shock which may cause the component to fail. Let the random variable (rv) $T_{i}$ denote lifetime of the $i$th component in the system which experiences a random shock at binging. Also suppose that $I_{i}$ denotes independent Bernoulli rvs, independent of the $X_i$’s, with $E(I_i)=p_i$, will be called shock parameter hereafter. Then, the random shock impacts the $i$th component ($I_{i}= 1$) with probability $p_i$ or doesn't impact the $i$th component ($I_{i}= 0$) with probability $1-p_i$. Hence, the rv $X_i =I_{i}T_{i}$ corresponds to the lifetime of the $i$th component in a system under shock. It is of interest to compare two such systems stochastically with respect to vector or matrix majorization.    
\\\hspace*{0.3in} Similar comparisons are carried out in the context of insurance where largest or smallest claim amounts in a portfolio of risks are compared stochastically. One may refer to Balakrishnan \emph{et al.}~\cite{ba2}, and Barmalzan\emph{et al.}~\cite{bar3} for more detail.      
\\\hspace*{0.3in} In reliability optimization and life testing experiments, many times the tests are censored or truncated. For example, failure of a device during the warranty period may not be counted or items may be replaced after a certain time under a replacement policy. Moreover, test conditions, cost or other constraints may lead many reliability systems to be bounded above. These situations result in a data set which is modeled by distributions with finite range i.e. with bounded support (cf. Chowdhury and Kundu~\cite{ch}). In this context, G$\acute{o}$mez et al.~\cite{go1} has proposed log-Lindley distribution which has a simple expression and flexible reliability properties as compared to the beta distribution. It exhibits bath-tub failure rates and has increasing generalized failure rate (IGFR). The distribution has useful applications in the context of inventory management, pricing and supply chain contracting problems where demand distribution is required to have the IGFR property (Ziya et al.~\cite{zia}, Lariviere and Porteus~\cite{lar1}, Lariviere~\cite{lar2}). Moreover, the distribution is found to be suitable for fitting rates and proportions data better than the existing distribution like beta. The paper by Chowdhury and Kundu~\cite{ch} has compared two parallel systems stochastically with log-Lindley distributed components assuming components fail with certainty. In this paper, we take the work a step forward and compare two $n$-components parallel systems having heterogeneous log-Lindley distributed components in terms of usual stochastic ordering when each of the components in the systems experiences a random shock. 
\\\hspace*{0.3in} As introduced by G$\acute{o}$mez et al.~\cite{go1}, the probability density function (pdf) and the cumulative distribution function (cdf) of the log-Lindley (LL) distribution, written as LL($\sigma,\lambda$) are given by
\begin{equation}\label{e0}
f(x;\sigma,\lambda)=\frac{\sigma^2}{1+\lambda\sigma}\left(\lambda-\log x\right) x^{\sigma-1};~0<x<1,~\lambda\geq 0,~\sigma>0,
\end{equation} 
and
\begin{equation}\label{e00}
F(x;\sigma,\lambda)=\frac{x^\sigma\left[1+\sigma\left(\lambda-\log x\right)\right]}{1+\lambda\sigma};~0<x<1,~\lambda\geq 0,~\sigma>0
\end{equation}
\hspace*{0.3in} The rest of the paper is organized as follows. In Section 2, we have given the required notations, definitions and some useful lemmas which are used throughout the paper. Results related to usual stochastic ordering between two parallel systems are derived in Section 3. Some concluding remarks are presented in Section 4. 
\\\hspace*{0.3 in} Throughout the paper, the word increasing (resp. decreasing) and nondecreasing (resp. nonincreasing) are used interchangeably, and $\mathbb{R_+}$ denotes the set of positive real numbers $\{x:0<x<\infty\}$. Moreover, $x>0$ denotes $0<x<1$ and correspondingly $\log x$ is negative throughout. We also write $a\stackrel{sign}{=}b$ to mean that $a$ and $b$ have the same sign and $h^{-1}$ denotes inverse of the function $h$. It is assumed that $0.\infty=0$. 
\section{Notations, Definitions and Preliminaries}
\hspace*{0.3 in} Let $X$ and $Y$ be two absolutely continuous random variables  with survival functions $\overline F_{X}\left(\cdot\right)$ and $\overline F_{Y}\left(\cdot\right)$ respectively.
\\\hspace*{0.3 in} In order to compare different order statistics, stochastic orders are used for fair and reasonable comparison.
Different kinds of stochastic orders are developed and studied in the literature. The following well known definitions may be obtained in Shaked and Shanthikumar~\cite{shak1}.
\begin{d1}\label{de1}
Let $X$ and $Y$ be two absolutely continuous rvs with respective supports $(l_X,u_X)$ and $(l_Y,u_Y)$,
where $u_X$ and $u_Y$ may be positive infinity, and $l_X$ and $l_Y$ may be negative infinity.
Then, $X$ is said to be smaller than $Y$ in
usual stochastic (st) order, denoted as $X\leq_{st}Y$, if $\bar F_X(t)\leq \bar F_Y(t)$ for all $t\in (-\infty,\infty).$ 
\end{d1}
\hspace*{0.3 in} It is well known that the results on different stochastic orders can be established on using majorization order(s). Let $I^n$ denote an $n$-dimensional Euclidean space where $I\subseteq\Re$. Further, let $\mathbf{x}=(x_1,x_2,\dots,x_n)\in I^n$ and $\mathbf{y}=(y_1,y_2,\dots,y_n)\in I^n$ be any two real vectors with $x_{(1)}\le x_{(2)}\le\cdots\le x_{(n)}$ being the increasing arrangements of the components of the vector $\mathbf{x}$. The following definitions on vector majorization may be found in Marshall \emph{et al.} \cite{Maol}.\\
\begin{d1}\label{de12}
\begin{enumerate}
\item [i)] The vector $\mathbf{x} $ is said to majorize the vector $\mathbf{y} $ (written as $\mathbf{x}\stackrel{m}{\succeq}\mathbf{y}$) if
\begin{equation*}
\sum_{i=1}^j x_{(i)}\le\sum_{i=1}^j y_{(i)},\;j=1,\;2,\;\ldots, n-1,\;\;and \;\;\sum_{i=1}^n x_{(i)}=\sum_{i=1}^n y_{(i)}.
\end{equation*} 
\item [ii)] The vector $\mathbf{x}$ is said to weakly supermajorize the vector $\mathbf{y}$ (written as $\mathbf{x}\stackrel{\rm w}{\succeq} \mathbf{y}$) if
\begin{equation*}
 \sum\limits_{i=1}^j x_{(i)}\leq \sum\limits_{i=1}^j y_{(i)}\quad \text{for}\;j=1,2,\dots,n.
\end{equation*}
\item [iii)] The vector $\mathbf{x}$ is said to weakly submajorize the vector $\mathbf{y}$
 (written as $\mathbf{x}\;{\succeq}_{\rm w} \;\mathbf{y}$) if
\begin{equation*}
  \sum\limits_{i=j}^n x_{(i)}\geq \sum\limits_{i=j}^n y_{(i)}\quad \text{for}\;j=1,2,\dots,n.
\end{equation*}
\end{enumerate}
\end{d1}
\hspace*{0.3 in} It is easy to show that $\mathbf{x}\stackrel{\rm m}{\succeq}\mathbf{y}\Rightarrow\mathbf{x}\stackrel{\rm w}{\succeq} \mathbf{y}$.
\begin{d1}\label{de2}
A function $\psi:I^n\rightarrow\Re$ is said to be Schur-convex (resp. Schur-concave) on $I^n$ if 
\begin{equation*}
\mathbf{x}\stackrel{m}{\succeq}\mathbf{y} \;\text{implies}\;\psi\left(\mathbf{x}\right)\ge (\text{resp. }\le)\;\psi\left(\mathbf{y}\right)\;for\;all\;\mathbf{x},\;\mathbf{y}\in I^n.
\end{equation*}
\end{d1}
The following definitions related to matrix majorization may be found in Marshall \emph{et al.} \cite{Maol}.
\begin{d1}\label{de131}
\begin{enumerate}
\item[i)] A square matrix $\Pi_n$, of order $n$, is said to be a permutation matrix if each row and column has a single entry as $1$, and all other entries as zero. 
\item[ii)] A square matrix $P=(p_{ij})$, of order $n$, is said to be doubly stochastic if $p_{ij}\geq 0$, for all $i,j = 1,...n,
\sum^{n}_{i=1}p_{ij}=1, j = 1,...,n$ and $\sum^{n}_{j=1}p_{ij}=1, i = 1,...,n.$ 
\item[iii)] A square matrix $T_n$, of order $n$, is said to be $T-$transform matrix if it has the form $$T_n=\lambda I_n+(1-\lambda)\Pi_n;~0\leq\lambda\leq 1,$$ where $I_n$ is the identity matrix and $\Pi_n$ is the permutation matrix that only interchanges two co-ordinates.
\end{enumerate}
\end{d1}
\begin{d1}\label{13}
Consider the $m\times n$ matrices $A=\{a_{ij}\}$ and $B=\{b_{ij}\}$ with rows $\mathbf{a_1},...,\mathbf{a_m}$ and
$\mathbf{b_1},..., \mathbf{b_m}$, respectively.
\begin{enumerate}
\item[i)] $A$ is said to be larger than $B$ in chain majorization, denoted by $A>>B$, if there exists a finite set of $n\times n$ $T-$transform matrices $T_1,..., T_k$ such that $B=AT_1T_2...T_k$.
\item[ii)] $A$ is said to majorize $B$, denoted by $A>B$, if $A=BP$, where the $n\times n$ matrix $P$ is doubly stochastic. Since a product of $T-$transforms is doubly stochastic, it follows that $A>>B \Rightarrow A>B.$
\item[iii)]  $A$ is said to be larger than the matrix $B$ in row majorization, denoted by $A>^{row}B$, if $\mathbf{a_i}\stackrel{m}{\succeq}\mathbf{b_i}$ for $i=1,...,m$. It is clear that $A>B \Rightarrow A>^{row}B.$ 
\item[iv)] $A$ is said to be larger than the matrix $B$ in row weakly majorization, denoted by $A>^{w}B$, if $\mathbf{a_i}\stackrel{w}{\succeq}\mathbf{b_i}$ for $i=1,...,m$. It is clear that $A>^{row}B \Rightarrow A>^{w}B.$
\end{enumerate}
\end{d1}
Thus it can be written that
$$A>>B\Rightarrow A>B\Rightarrow A>^{row}B\Rightarrow A>^w B.$$
\begin{n1}
Let us introduce the following notations.
\begin{enumerate}
\item[(i)] $\mathcal{D}_{+}=\left\{\left(x_{1},x_2,\ldots,x_{n}\right):x_{1}\geq x_2\geq\ldots\geq x_{n}> 0\right\}$.
\item[(ii)] $\mathcal{E}_{+}=\left\{\left(x_{1},x_2,\ldots,x_{n}\right):0< x_{1}\leq x_2\leq\ldots\leq x_{n}\right\}$.
\item[(iii)] $\mathcal{U}_{n}=\left\{\left(\mbox{\boldmath $x$},\mbox{\boldmath $y$}\right)=\begin{bmatrix}
    x_1 & x_2 &.&.&.&x_n \\
    y_1 & y_2 &.&.&.&y_n
  \end{bmatrix}
:(x_i-x_j)(y_i-y_j)\geq 0; i,j=1,2,\ldots n\right\}$.
\item[(iv)] $\mathcal{V}_{n}=\left\{\left(\mbox{\boldmath $x$},\mbox{\boldmath $y$}\right)=\begin{bmatrix}
    x_1 & x_2 &.&.&.&x_n \\
    y_1 & y_2 &.&.&.&y_n
  \end{bmatrix}
:(x_i-x_j)(y_i-y_j)\leq 0; i,j=1,2,\ldots n\right\}$.
\end{enumerate}
\end{n1}

Let us first introduce the following lemmas which will be used in the next section to prove the results. 

\begin{l1}\label{l6}
 $\left(\text{Lemma 3.1 of Kundu \emph{et al.}\cite{kun1}}\right)$~Let $\varphi:\mathcal{D_{+}}\rightarrow \mathbb{R}$ be a function, continuously differentiable on the interior of $\mathcal{D_{+}}$. Then, for $\mf{x},\mf{y}\in \mathcal{D_{+}}$,
 \begin{eqnarray*}
  \mf{x}\stackrel{m}\succeq\mf{y}\;\text{implies}\;\varphi(\mf{x})\geq (\text{resp.}\;\leq)\;\varphi(\mf{y})
 \end{eqnarray*}
if, and only if,
$$\varphi_{(k)}(\mf{z})\;\text{is decreasing (resp. increasing) in}\;k=1,2,\dots,n,$$
where $\varphi_{(k)}(\mf{z})=\partial\varphi(\mf{z})/\partial z_k$ denotes the partial derivative of $\varphi$ with respect to its $k$th argument.$\hfill\Box$
\end{l1}
\begin{l1}\label{l7}
 $\left(\text{Lemma 3.3 of Kundu \emph{et al.}\cite{kun1}}\right)\;$~Let $\varphi:\mathcal{E_{+}}\rightarrow \mathbb{R}$ be a function, continuously differentiable on the interior of $\mathcal{E_{+}}$.
 Then, for $\mf{x},\mf{y}\in \mathcal{E_{+}}$,
 \begin{eqnarray*}
  \mf{x}\stackrel{m}\succeq\mf{y}\;\text{implies}\;\varphi(\mf{x})\geq (\text{resp.}\;\leq)\;\varphi(\mf{y})
 \end{eqnarray*}
if, and only if,
$$\varphi_{(k)}(\mf{z})\;\text{is increasing (resp. decreasing) in}\;k=1,2,\dots,n,$$
where $\varphi_{(k)}(\mf{z})=\partial\varphi(\mf{z})/\partial z_k$ denotes the partial derivative of $\varphi$ with respect to its $k$-th argument.$\hfill\Box$
\end{l1}

\begin{l1}\label{l10}
$\left(\text{Theorem A.8 of Marshall \emph{et al.}~\cite{Maol}}\;p.p.~87\right)$~Let $S\subseteq \mathbb{R}^n$. Further, let $\varphi:S\rightarrow\mathbb{R}$ be a function. Then for $\mbox{{\bf x}}$, $\mbox{{\bf y}}\in S$,
$$\mbox{{\bf x}}\succeq_{w} \mbox{{\bf y}}\Longrightarrow\varphi\left(\mbox{{\bf x}}\right)\ge(resp. \le)\varphi\left(\mbox{{\bf y}}\right)$$
if, and if, $\varphi$ is both increasing (resp. decreasing) and Schur-convex (resp. Schur-concave) on $S$. Similarly,
$${\bf x}\stackrel{w}\succeq {\bf y}\Longrightarrow\varphi\left({\bf x}\right)\ge(resp. \le)\varphi\left({\bf y}\right)$$
if, and if, $\varphi$ is both decreasing (resp. increasing) and Schur-convex (resp. Schur-concave) on $S$.
\end{l1}

\section{Main Results}
\setcounter{equation}{0}
\hspace{0.3in} For $i=1,2,\ldots,n$, let $T_i$ (resp. $W_i$) be $n$ independent nonnegative rvs following LL distribution as given in (\ref{e0}). Assuming $X_i=T_iI_i$ and $Y_i=W_iI_i^*$, the cdf of $X_i$ and $Y_i$, for $t>0$, are given by 
$$F_{X_i}(t)=1-P(T_iI_i\geq t)=1-P(T_iI_i\geq t\mid I_i=1)P(I_i=1)=1-p_i\left(1-t^{\sigma_i}+\frac{\sigma_i t^{\sigma_i}\log t}{1+\lambda_i\sigma_i}\right)$$ and 
$$F_{Y_i}(t)=1-P(W_iI_i^*\geq t)=1-P(W_iI_i^*\geq t\mid I_i^*=1)P(I_i^*=1)=1-p^*_i\left(1-t^{\theta_i}+\frac{\theta_i t^{\theta_i}\log t}{1+\delta_i\theta_i}\right)$$ respectively, where $E(I_i)=p_i$ and $E(I_i^*)=p_i^*$. For $t=0$, $F_{X_i}(0)=1-p_i$ and $G_{Y_i}(0)=1-p_i^*$.\\
 \\
\hspace*{0.3 in} If $F_{X_{n:n}}\left(\cdot\right)\left(G_{Y_{n:n}}\left(\cdot\right)\right)$ and $\overline{F}_{X_{1:n}}\left(\cdot\right)\left(\overline{G}_{Y_{1:n}}\left(\cdot\right)\right)$ be the cdf and the survival function of $X_{n:n}(Y_{n:n})$ and $X_{1:n}(Y_{1:n})$ respectively, where $\mbox{\boldmath$\sigma$}=\left(\sigma_1,\sigma_2,\ldots,\sigma_n\right)$, $\mbox{\boldmath$\theta$}=\left(\theta_1,\theta_2, \ldots,\theta_n\right)$, $\mbox{\boldmath$\lambda$}=\left(\lambda_1,\lambda_2,\ldots,\lambda_n\right)$ and $\mbox{\boldmath$\delta$}=\left(\delta_1,\delta_2,\ldots,\delta_n\right)$, then from (\ref{e00}) it can be written that
\begin{equation}\label{e111}
F_{X_{n:n}}\left(x\right)=\prod_{i=1}^n F_{X_i}\left(x\right)=\prod_{i=1}^n \left[1-p_i\left(1-x^{\sigma_i}+\frac{\sigma_i x^{\sigma_i}\log x}{1+\lambda_i\sigma_i}\right)\right],
\end{equation}
and
\begin{equation*}
G_{Y_{n:n}}\left(x\right)=\prod_{i=1}^n F_{Y_i}\left(x\right)=\prod_{i=1}^n \left[1-p^*_i\left(1-x^{\theta_i}+\frac{\theta_i x^{\theta_i}\log x}{1+\delta_i\theta_i}\right)\right],
\end{equation*}
for $x>0$, and $F_{X_{n:n}}\left(0\right)=\prod_{i=1}^n\left(1-p_i\right)$, $G_{Y_{n:n}}\left(0\right)=\prod_{i=1}^n\left(1-p_i^*\right)$.\\
The first two theorems show that usual stochastic ordering holds between two parallel systems of heterogeneous components under random shocks for fixed $\sigma$. Theorem \ref{th1} guarantees that for parallel systems of components having independent LL
distributed lifetimes with common shape parameter vector and heterogeneous scale parameter vector, the majorized shock parameter vector
leads to larger systems lifetime (better system reliability) in the sense of the usual stochastic ordering.

\begin{t1}\label{th1}
For $i=1,2,\ldots, n$, let $T_i$ and $W_i$ be two sets of mutually independent random variables with $T_i\sim LL\left(\sigma,\lambda_i\right)$ and $W_i\sim LL\left(\sigma,\lambda_i\right)$. Further, suppose that $I_i~(I^{*}_i)$ be a set of independent Bernoulli rv, independent of $T_i$'s ($W_i$'s) with $E(I_i)=p_i~(E(I^{*}_i)=p^{*}_i), i=1,2,...,n.$ If $h:[0, 1]\rightarrow \Re+$ is a differentiable and strictly convex function, then
\begin{enumerate}
\item[i)] $\mbox{\boldmath $h(p)$}\succeq_w\mbox{\boldmath $h(p^{*})$}\;\text{implies}\; X_{n:n}\ge_{st}Y_{n:n}$ if $\mbox{\boldmath $\lambda$}\in \mathcal{D}_+ (\mathcal{E}_+)$, $\mbox{\boldmath $h(p)$}\in \mathcal{D}_+ (\mathcal{E}_+)$, and $h(u)$ is increasing in $u$. 
 \item[ii)] $\mbox{\boldmath $h(p)$}\stackrel{w}{\succeq}\mbox{\boldmath $h(p^{*})$}\;\text{implies}\; X_{n:n}\ge_{st}Y_{n:n}$ if $\mbox{\boldmath $\lambda$}\in \mathcal{D}_+ (\mathcal{E}_+)$, $\mbox{\boldmath $h(p)$}\in \mathcal{E}_+ (\mathcal{D}_+)$, and $h(u)$ is decreasing in $u$,
\end{enumerate}
where $\mbox{\boldmath $h(p)$}=\left(h\left(p_1\right),h\left(p_2\right),\ldots,h\left(p_n\right)\right)$. 
\end{t1}
{\bf Proof:} 
For $x>0$, in view of the expression (\ref{e111}), $$F_{X_{n:n}}\left(x\right)=\prod_{i=1}^n \left[1-h^{-1}(u_i)\left(1-x^{\sigma}+\frac{\sigma x^{\sigma}\log x}{1+\lambda_i\sigma}\right)\right]=\Psi(\mbox{\boldmath $u$)} (say),$$ 
where $h(p_i)=u_i$. Differentiating $\Psi(\bf u)$ partially, with respect to $u_i$, we get 
\begin{equation}\label{eq11}
\frac{\partial\Psi}{\partial u_i}=-\frac{dh^{-1}(u_i)}{du_i}\left(1-x^{\sigma}+\frac{\sigma x^{\sigma}\log x}{1+\lambda_i\sigma}\right)\prod^{n}_{k\neq i=1}\left[1-h^{-1}(u_k)\left(1-x^{\sigma}+\frac{\sigma x^{\sigma}\log x}{1+\lambda_i\sigma}\right)\right] \leq (\geq) 0,
\end{equation}
if $h(u)$ is increasing (decreasing) in $u$. Now,
\begin{equation}\label{e112}
\begin{split}
\frac{\partial\Psi}{\partial u_i}-\frac{\partial\Psi}{\partial u_j}&\stackrel{sign}{=}\prod^{n}_{k\neq i,j=1}\left[1-h^{-1}(u_k)w(x;\sigma,\lambda_k)\right]\left[\left\{1-h^{-1}(u_j)\left(w(x;\sigma,\lambda_j)\right)\right\}\left\{-\frac{dh^{-1}(u_i)}{du_i} w(x;\sigma,\lambda_i)\right\}\right.\\&\quad-\left.\left\{1-h^{-1}(u_i)w(x;\sigma,\lambda_i)\right\}\left\{-\frac{dh^{-1}(u_j)}{du_j} w(x;\sigma,\lambda_j)\right\}\right],
\end{split}
\end{equation}
where $w(x;\sigma,\lambda_i)=\left(1-x^{\sigma}+\frac{\sigma x^{\sigma}\log x}{1+\lambda_i\sigma}\right)$.\\
\hspace*{0.3 in} Now, two cases may arise:\\
$Case (i)$ For $i\leq j$, if $\lambda_i\geq (\leq)\lambda_j, u_i\geq (\leq)u_j$ and $h(u)$ is increasing and convex in $u$, then for all $\sigma\geq ((\leq)) 0$ and $0 \leq x\leq 1$ it can be written that 
$$1-h^{-1}(u_i)w(x;\sigma,\lambda_i)\leq (\geq) 1-h^{-1}(u_j)w(x;\sigma,\lambda_j).$$ 
Again, if $h(u)$ is convex in $u$, then $u_i\geq (\leq) u_j$ gives $\frac{dh^{-1}(u_i)}{du_i} \geq (\leq)\frac{dh^{-1}(u_j)}{du_j}$ which yields 
$$\frac{dh^{-1}(u_i)}{du_i}w(x;\sigma,\lambda_i)\geq 
(\leq)\frac{dh^{-1}(u_j)}{du_j}w(x;\sigma,\lambda_j).$$ 
Substituting the results in \ref{e112}, we get $\frac{d\Psi}{du_i}-\frac{d\Psi}{du_j}\leq (\geq) 0$. Thus by Lemma \ref{l6} it can be proved that $\Psi$ is Schur-concave in $\mbox{\boldmath $u$}.$  Thus the result is proved by Lemma \ref{l10}.\\
$Case (ii)$ For $i\leq j$, if $\lambda_i\geq (\leq)\lambda_j,$ $u_i\leq (\geq) u_j$ and $h(u)$ is decreasing and convex in $u$, then, for all $\sigma\geq 0$ and $0 \leq x\leq 1$ it can be written that $$1-h^{-1}(u_i)w(x;\sigma,\lambda_i)\leq (\geq) 1-h^{-1}(u_j)w(x;\sigma,\lambda_j).$$ 
As $h(u)$ is decreasing and convex $u$, $u_i\leq (\geq) u_j$ implies $\frac{dh^{-1}(u_i)}{du_i} \leq (\geq) \frac{dh^{-1}(u_j)}{du_j}$ which gives 
$$\frac{-dh^{-1}(u_i)}{du_i}w(x;\sigma,\lambda_i)\geq (\leq) \frac{-dh^{-1}(u_j)}{du_j}w(x;\sigma,\lambda_j).$$
Following the same argument as in $case (i)$, it can be shown that $\frac{d\Psi}{du_i}-\frac{d\Psi}{du_j}\geq (\leq)0$. Thus by Lemma \ref{l7} it can be proved that $\Psi$ is Schur-concave in $\mbox{\boldmath $u$}.$ So, the result follows from Lemma \ref{l10}.\\
Observing the fact that $w\left(x,\sigma,\lambda_i\right)=1$ when $x=0$, the theorem, for $x=0$, can be proved in the same line as above.   
This proves the result. $\hfill\Box$\\
For fixed shape parameter vector, the next theorem guarantees that parallel systems of components having independent LL distributed lifetimes heterogeneous shock parameter vector, the majorized scale parameter vector leads to smaller systems lifetime (worse system reliability) in the sense of the usual stochastic ordering.
\begin{t1}\label{th2}
For $i=1,2,\ldots, n$, let $T_i$ and $W_i$ be two sets of mutually independent random variables with $T_i\sim LL\left(\sigma,\lambda_i\right)$ and $W_i\sim LL\left(\sigma,\delta_i\right)$. Further, suppose that $I_i$ be a set of independent Bernoulli rvs, independent of $X_i$'s or $Y_i$'s with $E(I_i)=p_i,$ $i=1,2,...,n$ and $h:[0, 1]\rightarrow \Re+$ is a differentiable and strictly convex function. If
\begin{enumerate}
\item[i)] either $\mbox{\boldmath $\lambda$}\in \mathcal{D}_+$, $\mbox{\boldmath $h(p)$}\in \mathcal{D}_+(\mathcal{E}_+)$, and $h(u)$ is increasing (decreasing) in $u$,\\
\item[ii)] or $\mbox{\boldmath $\lambda$}\in \mathcal{E}_+$, $\mbox{\boldmath $h(p)$}\in \mathcal{E}_+(\mathcal{D}_+)$, and $h(u)$ is increasing (decreasing) in $u$,
\end{enumerate}
then $\mbox{\boldmath $v$}\stackrel{w}{\succeq}\mbox{\boldmath $v^{*}$}\;\text{implies}\; X_{n:n}\geq_{st}Y_{n:n},$ 
where $\mbox{\boldmath $v$}=\left(\frac{1}{1+\lambda_1\sigma},\frac{1}{1+\lambda_2\sigma},...,\frac{1}{1+\lambda_n\sigma}\right)$ and $\mbox{\boldmath $v^{*}$}=\left(\frac{1}{1+\delta_1\sigma},\frac{1}{1+\delta_2\sigma},...,\frac{1}{1+\delta_n\sigma}\right)$.
\end{t1}
{\bf Proof:} Assuming $v_i=\frac{1}{1+\lambda_i\sigma}$, (\ref{e111}) can be written as $$\Psi_1(\mbox{\boldmath $v$)}=\prod_{i=1}^n \left[1-h^{-1}(u_i)\left(1-x^{\sigma}+v_i\sigma x^{\sigma}\log x\right)\right].$$ 
Differentiating $\Psi_1(\mbox{\boldmath $v$})$ with respect to $v_i$, we get 
$$\frac{\partial\Psi_1}{\partial v_i}=\left(-h^{-1}(u_i)\sigma x^{\sigma}\log x\right)\prod^{n}_{k\neq i=1}\left[1-h^{-1}(u_k)\left(1-x^{\sigma}+v_k\sigma x^{\sigma}\log x\right)\right] \geq 0,$$ 
proving that $\Psi_1(\mbox{\boldmath $v$})$ is increasing in each $v_i$. Again, it can be easily shown that \\
\begin{equation*}
\frac{d\Psi_1}{dv_i}-\frac{d\Psi_1}{dv_j}=\frac{h^{-1}(u_i)}{1-h^{-1}(u_i)\left(1-x^{\sigma}+v_i\sigma x^{\sigma}\log x\right)}-\frac{h^{-1}(u_j)}{1-h^{-1}(u_j)\left(1-x^{\sigma}+v_j\sigma x^{\sigma}\log x\right)}.
\end{equation*}
Now, for $i\leq j$, $\lambda_i\geq\lambda_j$ implies $v_i\leq v_j,$ which in turn implies that 
$$1-x^{\sigma}+v_i\sigma x^{\sigma}\log x\geq 1-x^{\sigma}+v_j\sigma x^{\sigma}\log x.$$
Again, $u_i\geq u_j$ ($u_i\leq u_j$) and $h(u)$ is increasing (decreasing) in $u$ imply that $$1-h^{-1}(u_i)\left(1-x^{\sigma}+v_i\sigma x^{\sigma}\log x\right)\leq 1-h^{-1}(u_j)\left(1-x^{\sigma}+v_i\sigma x^{\sigma}\log x\right)$$ 
which eventually gives
$$\frac{h^{-1}(u_i)}{1-h^{-1}(u_i)\left(1-x^{\sigma}+v_i\sigma x^{\sigma}\log x\right)}\geq \frac{h^{-1}(u_j)}{1-h^{-1}(u_j)\left(1-x^{\sigma}+v_j\sigma x^{\sigma}\log x\right)} i.e. \frac{d\Psi_1}{dv_i}-\frac{d\Psi_1}{dv_j}\geq 0$$. \\
Therefore, by Lemma \ref{l7}, $\Psi_1$ Schur-concave in $\mbox{\boldmath $v$}$. Thus by Lemma \ref{l10} the result is proved.\\
For $\mbox{\boldmath $\lambda$}\in \mathcal{E}_+$, $\mbox{\boldmath $h(p)$}\in \mathcal{E}_+(\mathcal{D}_+)$, and $h(u)$ is increasing (decreasing) in $u$, then the theorem can be proved in similar way.$\hfill\Box$\\
Now the question arises$-$ what will happen if both the scale and shock parameter vectors i.e. the matrix of scale and shock parameters of one system  majorizes the other when the shape parameter vector remains constant? The theorem given below answers that the majorized matrix of the parameters leads to better system reliability. Combining Theorem \ref{th1}~(ii) and Theorem \ref{th2} (bracketed portion), the following theorem can be obtained.

\begin{t1}\label{th3}
For $i=1,2,\ldots, n$, let $T_i$ and $W_i$ be two sets of mutually independent random variables with $T_i\sim LL\left(\sigma,\lambda_i\right)$ and $W_i\sim LL\left(\sigma,\delta_i\right)$. Further, suppose that $I_i~(I^{*}_i)$ be a set of independent Bernoulli rvs, independent of $X_i$'s ($Y_i$'s) with $E(I_i)=p_i~(E(I^{*}_i)=p^{*}_i),$ $i=1,2,...,n$ and $h:[0, 1]\rightarrow \Re+$ is a differentiable and strictly decreasing and convex function. If $(\mbox{\boldmath $v$},\mbox{\boldmath $h(p)$})\in \mathcal{U}_{n}$, and $(\mbox{\boldmath $v^*$},\mbox{\boldmath $h(p^*)$})\in \mathcal{U}_{n}$, then
$$\begin{bmatrix}
    \mbox{\boldmath $h(p)$} \\
   \mbox{\boldmath $v$}
  \end{bmatrix}>^{w}\begin{bmatrix}
    \mbox{\boldmath $h(p^*)$} \\
   \mbox{\boldmath $v^*$}
  \end{bmatrix}\;\text{implies}\; X_{n:n}\geq_{st}Y_{n:n}.$$ \\ 
where$~$$\mbox{\boldmath $v$}=\left(\frac{1}{1+\lambda_1\sigma},\frac{1}{1+\lambda_2\sigma},...,\frac{1}{1+\lambda_n\sigma}\right)$ and $\mbox{\boldmath $v^{*}$}=\left(\frac{1}{1+\beta_1\sigma},\frac{1}{1+\beta_2\sigma},...,\frac{1}{1+\beta_n\sigma}\right)$.
\end{t1}
The counterexample, given below justifies the above theorem.
\begin{e1}
For $\sigma=0.5$ and $i=1,2,3$, let $T_i\sim LL\left(\sigma, \lambda_i\right)$ and $W_i\sim LL\left(\sigma, \beta_i\right)$ be two sets of mutually independent random variables. Let $\left(v_1,v_2,v_3\right)=\left(0.4, 0.4,0.1\right)$ and $\left(v_1^*,v_2^*,v_3^*\right)=\left(0.5, 0.4,0.2\right)$, where $v_i=\frac{1}{1+\lambda_i\sigma}$ and $v_i^*=\frac{1}{1+\beta_i\sigma}$. Now if $ $\mbox{\boldmath $h(p)$}$=\left(2,2,1\right)$  and $ $\mbox{\boldmath $h(p^*)$}$=\left(3,2,1\right)$  are taken, where $h(u)=-\log(u)$, then it can be written that 
$\begin{bmatrix}
    \mbox{\boldmath $h(p)$} \\
   \mbox{\boldmath $v$}
  \end{bmatrix}>^{w}\begin{bmatrix}
    \mbox{\boldmath $h(p^*)$} \\
   \mbox{\boldmath $v^*$}
  \end{bmatrix}$, where $(\mbox{\boldmath $v$},\mbox{\boldmath $h(p)$})\in \mathcal{U}_{n}$, and $(\mbox{\boldmath $v^*$},\mbox{\boldmath $h(p^*)$})\in \mathcal{U}_{n}$. The figure given below also shows that, for all $0\leq x\leq 1$, $F_{X_{3:3}}(x)\leq F_{Y_{3:3}}(x)$ giving that $X_{3:3}\geq_{st}Y_{3:3}$.
  \begin{figure}[ht]
\centering
\includegraphics[height=6.5 cm]{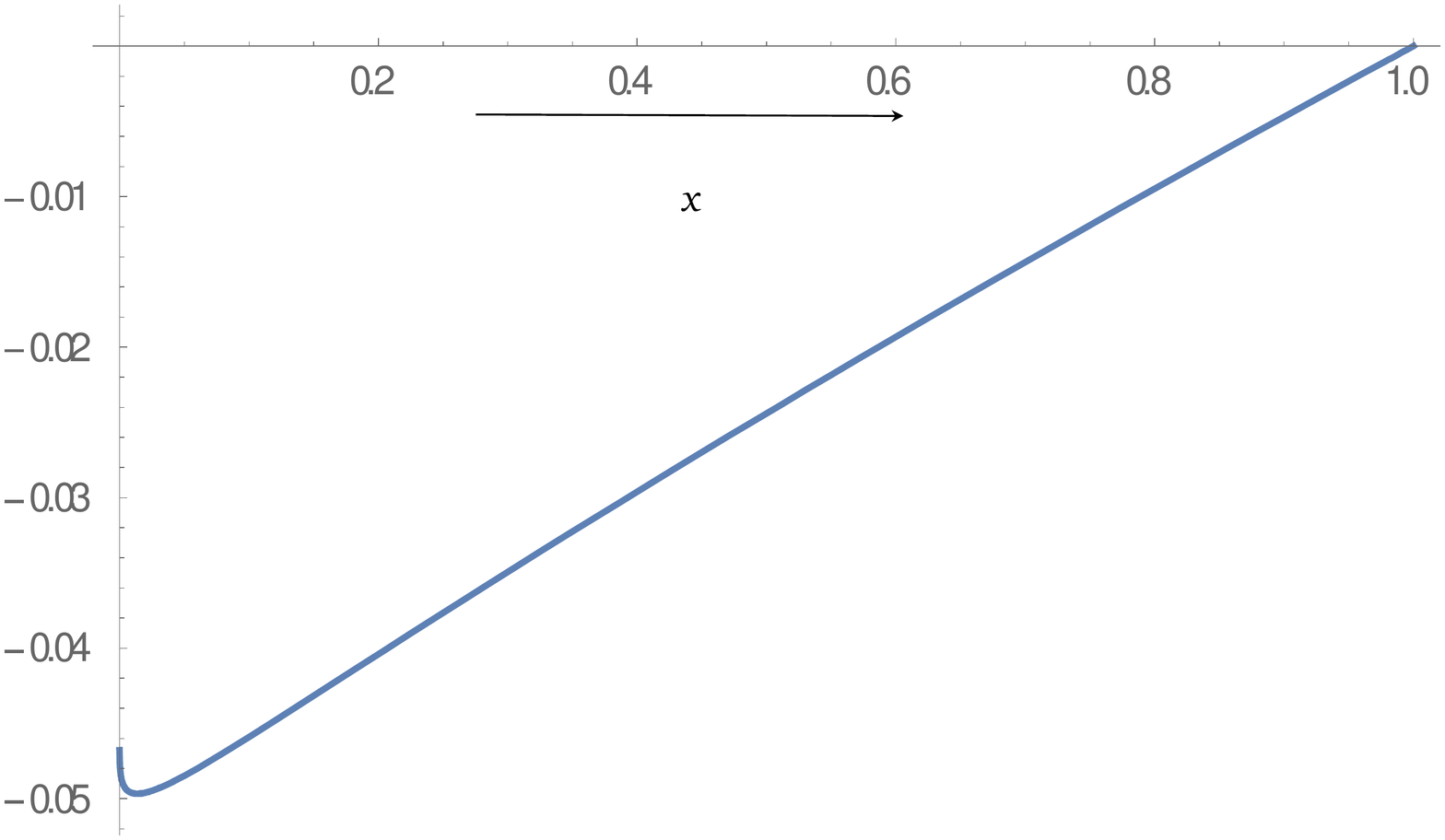}
\caption{\label{fig1}Graph of $F_{X_{3:3}}(x)-F_{Y_{3:3}}(x)$}
\end{figure}
\end{e1}
Thus, from the previous theorem the next theorem can be concluded.
\begin{t1}\label{th6}
For $i=1,2,\ldots, n$, let $T_i$ and $W_i$ be two sets of mutually independent random variables with $T_i\sim LL\left(\sigma,\lambda_i\right)$ and $W_i\sim LL\left(\sigma,\delta_i\right)$. Further, suppose that $I_i~(I^{*}_i)$ be a set of independent Bernoulli rvs, independent of $X_i$'s ($Y_i$'s) with $E(I_i)=p_i~(E(I^{*}_i)=p^{*}_i),$ $i=1,2,...,n.$ If $h:[0, 1]\rightarrow \Re+$ is a differentiable, strictly decreasing and and convex function, and also $(\mbox{\boldmath $v$},\mbox{\boldmath $h(p)$})$, $(\mbox{\boldmath $v^*$},\mbox{\boldmath $h(p^*)$})\in \mathcal{U}_{n}$, then
$$\begin{bmatrix}
    \mbox{\boldmath $h(p)$} \\
   \mbox{\boldmath $v$}
  \end{bmatrix}>>\begin{bmatrix}
    \mbox{\boldmath $h(p^*)$} \\
   \mbox{\boldmath $v^*$}
  \end{bmatrix}\;\text{implies}\; X_{n:n}\geq_{st}Y_{n:n}.$$
\end{t1}
Now the question arises$-$ can we have similar result as in Theorem \ref{th1} when the shape parameter vectors are heterogeneous and the scale parameter vector is constant? The theorem given below answers that the majorized shock parameter vector leads to larger systems lifetime (better system reliability) in the sense of the usual stochastic ordering in this case also.
\begin{t1}\label{th7}
For $i=1,2,\ldots, n$, let $T_i$ and $W_i$ be two sets of mutually independent random variables with $T_i\sim LL\left(\sigma_i,\lambda\right)$ and $W_i\sim LL\left(\sigma_i,\lambda\right)$. Further, suppose that $I_i~(I^{*}_i)$ be a set of independent Bernoulli rv, independent of $X_i$'s ($Y_i$'s) with $E(I_i)=p_i~(E(I^{*}_i)=p^{*}_i), i=1,2,...,n.$ If $h:[0, 1]\rightarrow \Re+$ is a differentiable and strictly convex function, then
\begin{enumerate}
\item[i)] $\mbox{\boldmath $h(p)$}\succeq_w\mbox{\boldmath $h(p^{*})$}\;\text{implies}\; X_{n:n}\ge_{st}Y_{n:n}$ if $\mbox{\boldmath $\sigma$}\in \mathcal{D}_+$, $\mbox{\boldmath $h(p)$}\in \mathcal{D}_+$, and $h(u)$ is increasing in $u$. 
 \item[ii)] $\mbox{\boldmath $h(p)$}\stackrel{w}{\succeq}\mbox{\boldmath $h(p^{*})$}\;\text{implies}\; X_{n:n}\ge_{st}Y_{n:n}$ if $\mbox{\boldmath $\lambda$}\in \mathcal{D}_+$, $\mbox{\boldmath $h(p)$}\in \mathcal{E}_+$, and $h(u)$ is decreasing in $u$.
\end{enumerate}
\end{t1}
{\bf Proof:} 
For $x>0$, from \ref{e111}, let us assume $\Psi_2(\mbox{\boldmath $u$})=\prod_{i=1}^n \left[1-h^{-1}(u_i)\left(1-x^{\sigma_i}+\frac{\sigma_i x^{\sigma_i}\log x}{1+\lambda\sigma_i}\right)\right]$. If $h(u)$ is increasing (decreasing) in $u$, then from (\ref{eq11}) it can be shown that $\frac{\partial\Psi}{\partial u_i} \leq (\geq) 0.$  Moreover, following (\ref{eq11}), we get  
\begin{equation}\label{e115}
\begin{split}
\frac{\partial\Psi_2}{\partial u_i}-\frac{d\Psi_2}{du_j}&\stackrel{sign}{=}\prod^{n}_{k\neq i,j=1}\left[1-h^{-1}(u_k)w_1(x;\sigma_k,\lambda)\right]\left[\left\{1-h^{-1}(u_j)w_1(x;\sigma_j,\lambda)\right\}\left\{-\frac{dh^{-1}(u_i)}{du_i} w_1(x;\sigma_i,\lambda)\right\}\right.\\&\quad-\left.\left\{1-h^{-1}(u_i)w(x;\sigma_i,\lambda)\right\}\left\{-\frac{dh^{-1}(u_j)}{du_j} w(x;\sigma_j,\lambda)\right\}\right],
\end{split}
\end{equation}
where $w_1(x;\sigma_i,\lambda)=1-x^{\sigma_i}+\frac{\sigma_i x^{\sigma_i}\log x}{1+\lambda\sigma_i}$.\\
\hspace*{0.3 in} Now, two cases may arise:\\
$Case (i)$ For $i\leq j$, let $\sigma_i\geq\sigma_j, u_i\geq u_j,$ and $h(u)$ is increasing and convex in $u$. Now, differentiating $w_1(x;\sigma_i,\lambda)$ with respect to $\sigma_i$, we get 
$$\frac{d w_1}{d\sigma_i}=x^{\sigma_i}\log x\left[-1+\frac{1}{(1+\lambda\sigma_i)^2}+\frac{\sigma_i\log x}{1+\lambda\sigma_i}\right]\geq 0,$$ implying that $w_1(x;\sigma_i,\lambda)$ is increasing in each $\sigma_i$. Moreover, $w_1(x;\sigma_i,\lambda)=0$ at $\sigma_i=0$.Thus $w_1(x;\sigma_i,\lambda)\geq 0$ for all $\sigma_i; i=1,2,...,n.$ \\
As $\sigma_i\geq\sigma_j,$ $u_i\geq u_j$ and $h(u)$ is increasing in $u$, it is obvious that $$1-h^{-1}(u_i)\left(w_1(x;\sigma_i,\lambda)\right)\leq 1-h^{-1}(u_j)\left(w_1(x;\sigma_j,\lambda)\right).$$ Also, as $h(u)$ is convex in $u$, then $u_i\geq u_j$ implies $\frac{dh^{-1}(u_i)}{du_i} \geq \frac{dh^{-1}(u_j)}{du_j}$, which in turn gives 
$$\frac{dh^{-1}(u_i)}{du_i}w_1(x;\sigma_i,\lambda)\geq \frac{dh^{-1}(u_j)}{du_j}w_1(x;\sigma_j,\lambda).$$ 
Substituting the results in (\ref{e115}), we get $\frac{d\Psi_2}{du_i}-\frac{d\Psi_2}{du_j}\leq 0$, which, by Lemma \ref{l6}, gives that $\Psi_2$ is Schur-concave in $\mbox{\boldmath $u$}.$  Thus, by Lemma \ref{l10} the result is proved.\\
$Case (ii)$ Again, for $i\leq j$, let $\sigma_i\geq\sigma_j, u_i\leq u_j,$ and $h(u)$ is decreasing and convex in $u$. Then, following the same argument and similar line of proof as in $Case (i)$, it can be proved that $\frac{d\Psi_2}{du_i}-\frac{d\Psi_2}{du_j}\geq 0$ proving that $\Psi_2$ is Schur-concave in $\mbox{\boldmath $u$}$, by Lemma \ref{l7}. Thus, by Lemma \ref{l10} the result is proved.\\   
The theorem, for $x=0$, can be proved in the similar line as above. This proves the result. $\hfill\Box$\\
For fixed $\lambda$ and equal $\mbox{\boldmath $\sigma$}$, although the previous theorem shows that there exists stochastic ordering between $X_{n:n}$ and $Y_{n:n}$ when  $\mbox{\boldmath $h(p)$}$ and  $\mbox{\boldmath $h(p^*)$}$ are odered in the sense of majorization, the next counterexample shows that no such ordering exists between $X_{n:n}$ and $Y_{n:n}$ when $\mbox{\boldmath $\sigma$}$ and $\mbox{\boldmath $\sigma^*$}$ are ordered in the sense of majorization, keeping $\mbox{\boldmath $\lambda$}$ and $\mbox{\boldmath $h(p)$}$ as equal for both the systems.
\begin{e1}
Let $X_i\sim LL\left(\sigma_i, \lambda_i\right)$ and $Y_i\sim\left(\sigma_i^*, \lambda_i \right), i=1,2,3$. Let $\mbox{\boldmath $\lambda$}=(0.5,0.5,0.5)$, $\left(\sigma_1, \sigma_2, \sigma_3\right)=\left(3,2,1\right)\in \mathcal{D}_+$ and $\left(\sigma_1^*, \sigma_2^*, \sigma_3^*\right)=\left(2,2,2\right)\in \mathcal{D}_+$, giving $\mbox{\boldmath $\sigma$}\stackrel{m}{\succeq}\mbox{\boldmath $\sigma^*$}$. Now, if $ $\mbox{\boldmath $h(p)$}$=\left(1,2,3\right)\in \mathcal{E}_+$ is taken, where $h(u)=-\log u$, then Figure \ref{fig1}(a) shows that there exists no stochastic ordering between $X_{3:3}$ and $Y_{3:3}$.\\
Again, for the same $h(u)$, if $ $\mbox{\boldmath $h(p)$}$=\left(0.03,0.02,0.01\right)\in \mathcal{D}_+$, $\left(\sigma_1, \sigma_2, \sigma_3\right)=\left(3,2,1\right)\in \mathcal{D}_+$ and $\left(\sigma_1^*, \sigma_2^*, \sigma_3^*\right)=\left(2.6,2.4,1\right)\in \mathcal{D}_+$ are taken, giving $\mbox{\boldmath $\sigma$}\stackrel{m}{\succeq}\mbox{\boldmath $\sigma^*$}$, then for $\mbox{\boldmath $\lambda$}=(0.5,0.5,0.5)$, Figure \ref{fig1}(b) also shows that there exists no stochastic ordering between $X_{3:3}$ and $Y_{3:3}$.
\begin{figure}[ht]
\centering
\begin{minipage}[b]{0.48\linewidth}
\includegraphics[height=6.5 cm]{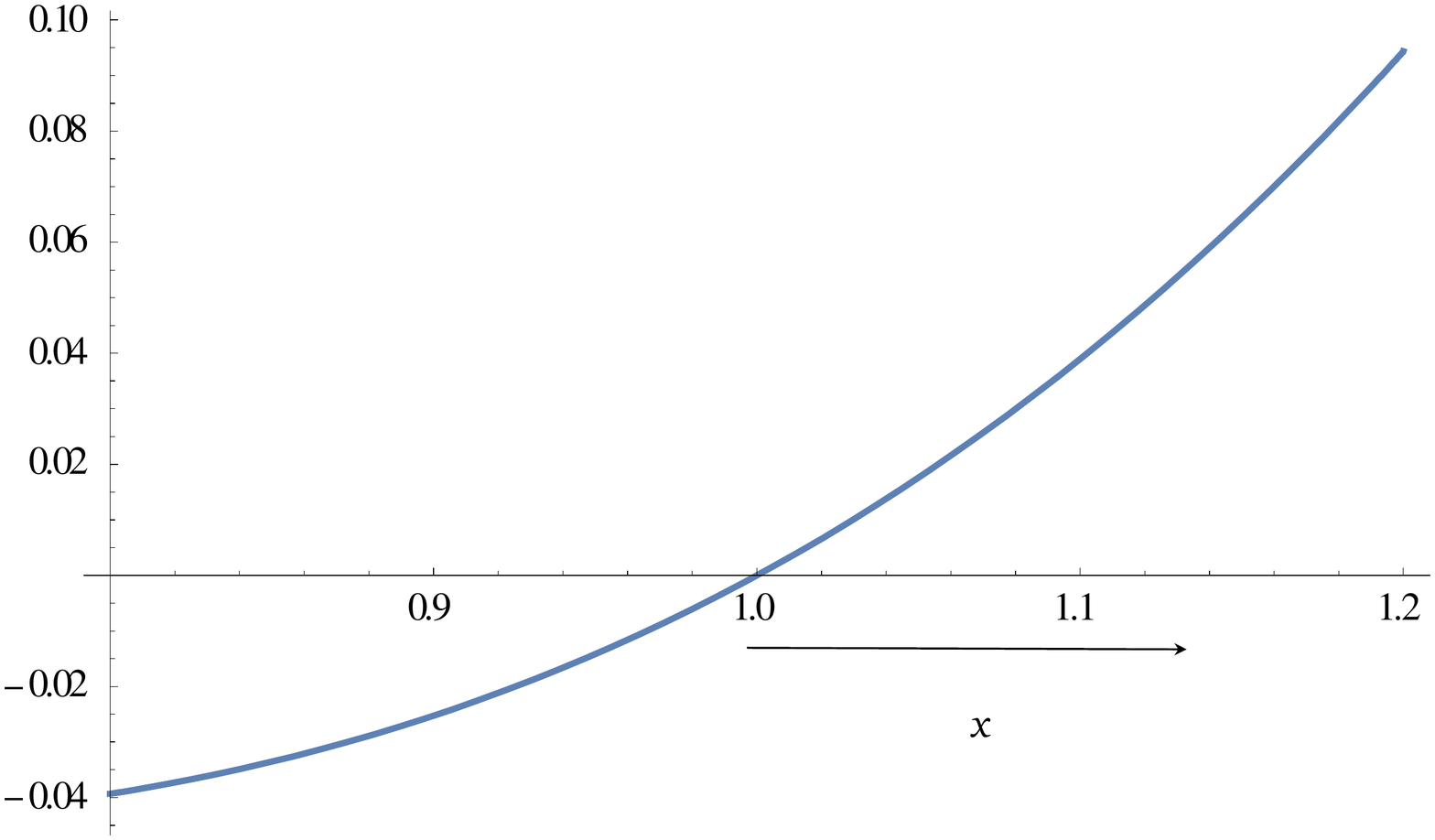}
\centering{(a) For $\mbox{\boldmath $\lambda$}, \mbox{\boldmath $\sigma$}, \mbox{\boldmath $\sigma^*$}\in\mathcal{D^+} \;and\; \mbox{\boldmath $h(p)$}\in \mathcal{E^+}$}
\end{minipage}
\quad
\begin{minipage}[b]{0.48\linewidth}
\includegraphics[height=6.5 cm]{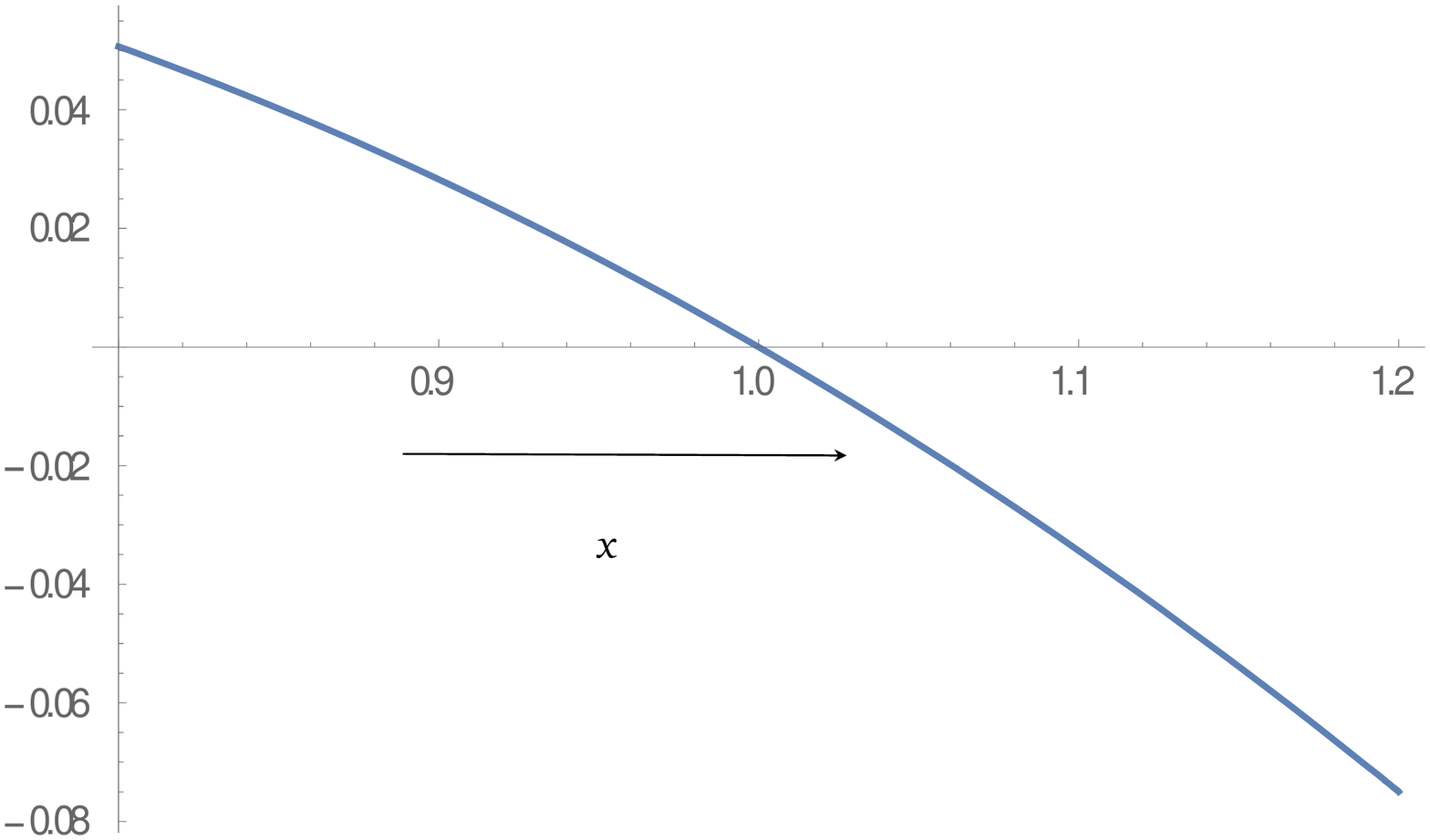}
\centering{(b) For $\mbox{\boldmath $\lambda$}, \mbox{\boldmath $\sigma$}, \mbox{\boldmath $\sigma^*$} \;and\; \mbox{\boldmath $h(p)$}\in \mathcal{D^+}$}
\end{minipage}
\caption{\label{fig1}Graph of $\overline F_{X_{3:3}}(x)-\overline F_{Y_{3:3}}(x)$}
\end{figure}
\end{e1}
\section{Concluding Remarks}
\hspace*{0.3 in} It is known that order statistics play an important role in reliability optimization and life testing experiments. Parallel systems being one of the building blocks of many complex coherent systems are required to be compared stochastically. Such comparisons are generally carried out with the assumption that the components of the system fail with certainty. In practice, the components may experience random shocks which eventually doesn't guarantee its failure. This paper compares the lifetimes of two parallel systems having heterogeneous log-Lindley distributed components under random shocks. It is proved that for two parallel systems with common shape parameter vector, the majorized matrix of the scale and shock parameters leads to better system reliability. It is also shown through counterexamples that no such results exist when the matrix of shape and shock parameters of one system majorizes the same of the other.   


\end{document}